\documentclass[12pt,a4paper,twoside]{article}

\usepackage[applemac]{inputenc}
\usepackage[english]{babel}
\usepackage[T1]{fontenc}
\usepackage{lmodern}
\usepackage{geometry} 
\usepackage[intlimits,sumlimits]{amsmath}
\usepackage{amssymb}
\usepackage[amsthm,thmmarks]{ntheorem}
\usepackage{mathrsfs}
\usepackage{bm}
\usepackage{xpatch}
\usepackage{dsfont}
\usepackage{siunitx} 
\usepackage{scalerel,stackengine}
\usepackage{aligned-overset}
\usepackage{algorithm}
\usepackage{multicol}
\usepackage{booktabs}

\usepackage{makecell}

\usepackage[backend=bibtex, style=authoryear, firstinits=true, maxcitenames=2, sorting=nty, isbn=false, dashed=false]{biblatex}
\DeclareNameAlias{sortname}{last-first}
\DeclareFieldFormat*{title}{#1}
\DeclareFieldFormat*{edition}{#1 edition}
\DeclareDelimFormat[cite,parencite]{nameyeardelim}{\textcolor{black}{\addcomma\space}}

\DeclareFieldFormat{citehyperref}{%
   \DeclareFieldAlias{bibhyperref}{noformat}
   \bibhyperref{#1}}

\savebibmacro{cite}

\renewbibmacro*{cite}{%
   \printtext[citehyperref]{%
      \restorebibmacro{cite}%
      \usebibmacro{cite}}}
\usepackage{xpatch}
\xpatchbibmacro{textcite}
  {\printnames{labelname}}
  {\printtext[bibhyperref]{\printnames{labelname}}}
  {}{}

\usepackage{caption}
\usepackage{subcaption}
\usepackage{paralist}
\usepackage{graphicx}
\usepackage{colortbl}
\usepackage[percent]{overpic}
\usepackage{mathtools}
\usepackage[usenames,dvipsnames,svgnames,table]{xcolor}
\usepackage{pgf,tikz}
\usetikzlibrary{patterns}
\usetikzlibrary{quotes,babel,angles}
\usetikzlibrary{arrows}
\usepackage{rotating}
\usepackage{pgfplots}
\usepackage{wrapfig}
\usepackage[most]{tcolorbox}
\usepackage[compact]{titlesec}
\usepackage{fancyhdr}

\definecolor{UHwhite}{RGB}{255,255,255}
\definecolor{UHorange}{RGB}{255,128,0}
\definecolor{UHred}{RGB}{202,53,56}
\definecolor{UHgruen}{RGB}{161,176,45}
\definecolor{UHdarkblue}{RGB}{6,68,107}
\definecolor{UHblue}{RGB}{24,105,183}
\definecolor{UHmidblue}{RGB}{17,119,182}
\definecolor{UHcyan}{RGB}{87,168,211}
\definecolor{UHbrightblue}{RGB}{160,206,234}
\definecolor{UHdarkgrey}{RGB}{69,69,69}
\definecolor{UHgrey}{RGB}{148,148,148}
\definecolor{UHbrightgrey}{RGB}{215,215,215}
\definecolor{UHbeige}{RGB}{235,230,215}

\definecolor{myred}{RGB}{178,34,34}
\definecolor{mygrey}{RGB}{69,69,69}

\numberwithin{equation}{section}
\definecolor{blau}{RGB}{24,105,183}
\usepackage[colorlinks=False,
pdfpagelabels,
pdfstartview = FitH,
bookmarksopen = true,
bookmarksnumbered = true,
colorlinks   = true,
linkcolor = black,
citecolor = UHblue,
plainpages = false,
hypertexnames = false,
linkbordercolor = UHblue,
urlcolor     = UHblue,
citebordercolor=white,
urlbordercolor=white]{hyperref}
\usepackage{orcidlink}

\usepackage{sectsty}
\usepackage{booktabs}

\usepackage[justification=centering]{caption}
\usepackage{dcolumn}
\usepackage{framed}
\newcolumntype{2}{D{.}{}{3.0}}
\makeatletter 
\renewcommand*{\@pnumwidth}{3em}      
\makeatother
\makeatletter
\renewcommand{\p@envcount}{\thesubsection.}
\makeatother

\newcommand{\N}{\mathbb{N}}

\newcommand{\R}{\mathbb{R}}

\newcommand{\eps}{\varepsilon}

\newcommand{\spn}[1]{\mathrm{span}\kern-0.4ex\left(#1\right)}
\newcommand{\supp}[1]{\mathrm{supp}\kern-0.4ex\left(#1\right)}
\newcommand{\bild}[1]{\mathrm{Bild}\kern-0.4ex\left( #1\right)}
\newcommand{\krn}[1]{\mathrm{Kern}\kern-0.4ex\left( #1\right)}
\newcommand{\rang}[1]{\mathrm{rang}\kern-0.4ex\left( #1\right)}
\newcommand{\spur}[1]{\mathrm{Spur}\kern-0.4ex\left( #1\right)}

\newcommand{\ggt}[1]{\mathrm{ggT}\kern-0.4ex\left( #1\right)}
\newcommand{\kgv}[1]{\mathrm{kgV}\kern-0.4ex\left( #1\right)}

\newcommand{\grpGL}[1]{\mathrm{GL}\kern-0.4ex\left( #1\right)}
\newcommand{\grpSL}[1]{\mathrm{SL}\kern-0.4ex\left( #1\right)}
\newcommand{\grpO}[1]{\mathrm{O}\kern-0.4ex\left( #1\right)}
\newcommand{\grpSO}[1]{\mathrm{SO}\kern-0.4ex\left( #1\right)}
\newcommand{\grpU}[1]{\mathrm{U}\kern-0.4ex\left( #1\right)}
\newcommand{\grpSU}[1]{\mathrm{SU}\kern-0.4ex\left( #1\right)}

\renewcommand{\P}[1]{\mathbb{P}\kern-0.4ex\left( #1\right)}
\newcommand{\E}[1]{\mathbb{E}\kern-0.4ex\left( #1\right)}

\makeatletter
\renewenvironment{abstract}{%
    \if@twocolumn
      \section*{\abstractname}%
    \else 
      \begin{center}%
        {\bfseries \abstractname\vspace{\z@}}%
      \end{center}%
      \quotation
    \fi}
    {\if@twocolumn\else\endquotation\fi}
\makeatother

\begin{document}
\sloppy


\title{
\vspace*{-1cm} 
\fontsize{24}{30} \selectfont \textbf{Time Series based\\ Ensemble Model Output Statistics\\ for Temperature Forecasts Postprocessing}}

\author{David Jobst\,\orcidlink{0000-0002-2014-3530}\thanks{Corresponding author, University of Hildesheim, Institute of Mathematics and Applied Informatics, Samelsonplatz 1, 31141 Hildesheim, Germany, \texttt{\href{mailto:jobstd@uni-hildesheim.de}{jobstd@uni-hildesheim.de}}},\and  Annette M\"oller\,\orcidlink{0000-0001-9386-1691}\thanks{Bielefeld University, Faculty of Business Administration and Economics, Universit\"atsstra{\ss}e 25, 33615 Bielefeld, Germany, \texttt{\href{mailto:annette.moeller@uni-bielefeld.de}{annette.moeller@uni-bielefeld.de}}} 
\and J\"urgen Gro{\ss}\,\orcidlink{0000-0002-3861-4708}
\thanks{University of Hildesheim, Institute of Mathematics and Applied Informatics, Samelsonplatz 1, 31141 Hildesheim, Germany, \texttt{\href{mailto:juergen.gross@uni-hildesheim.de}{juergen.gross@uni-hildesheim.de}}}}
\maketitle
\thispagestyle{empty}

\begin{abstract}
Nowadays, weather prediction is based on numerical weather prediction (NWP) models to produce an ensemble of forecasts. Despite of large improvements over the last few decades, they still tend to exhibit systematic bias and dispersion errors. Consequently, these forecasts may be improved by statistical postprocessing. This work proposes an extension of the ensemble model output statistics (EMOS) method in a time series framework. Besides of taking account of seasonality and trend in the location and scale parameter of the predictive distribution, the autoregressive process in the mean forecast errors or the standardized forecast errors is considered. The  models can be further extended by allowing generalized autoregressive conditional heteroscedasticity (GARCH). Last but not least, it is outlined how to use these models for arbitrary forecast horizons. To illustrate the performance of the suggested EMOS models in time series fashion, we present a case study for the postprocessing of 2\,\si{\meter} surface temperature forecasts using five different lead times and a set of observation stations in Germany. The results indicate that the time series EMOS extensions are able to significantly outperform the benchmark EMOS and autoregressive adjusted EMOS (AR-EMOS) in most of the lead time-station cases. To complement this article, our method is accompanied by an \texttt{R}-package called \texttt{tsEMOS}. 
\end{abstract}

\newpage

\section{Introduction}

Weather prediction is usually based on  numerical weather prediction (NWP) models, which are run multiple times with different model and/or initial and boundary conditions \parencites{Gneiting2005, Leutbecher2008}. The output is a forecast ensemble which allows for a probabilistic forecast to quantify forecast uncertainty \parencite{Palmer2002}.

In practice the NWP ensemble forecasts suffer from systematic errors, such as systematic bias and underdispersion and thus may benefit from statistical postprocessing based on past data to improve calibration and forecast skill. A well established and widely used postprocessing model is the so-called Ensemble Model Output Statistics (EMOS, \cites{Gneiting2005}). This method allows to provide a full predictive distribution where summary statistics of the forecast ensemble, such as, e.g., the mean or standard deviation are linked to the distribution parameters. The original EMOS was developed for Gaussian distributed weather quantities, as e.g., temperature or air pressure, and later extended for other weather quantities (see, e.g., \cites{Schefzik2013, Gneiting2014a, Hemri2014}).

Although the use of time series models for weather forecasting is quite common (see, e.g., \textcites{Tol1996, Cao2004, Campbell2005} for temperature time series models and \textcite{Brown1984, Tol1997,  Benth2010} for wind speed time series models) these approaches were only finding their way into the context of ensemble postprocessing when \textcite{Moeller2016} proposed a time series motivated model, the autoregressive adjusted EMOS (AR-EMOS), for postprocessing of 2\,\si{\meter} surface temperature forecasts. To account for possible autocorrelations in the one-step-ahead forecasts errors, they construct a predictive Gaussian distribution based on an autoregressive (AR)-adjusted forecast ensemble. Later, \textcite{Moeller2019} extended AR-EMOS, where they allow to vary the scale parameter of the Gaussian distribution by a convex combination of the empirical ensemble spread of the AR-adjusted forecast ensemble and the square root of mean variances obtained from the autoregressive process for arbitrary lead times.
In a benchmark study by \textcite{Demaeyer2023}, AR-EMOS shows comparable results to well-known as well as machine-learning based postprocessing techniques. Recently, \textcite{Tolomei2022} proposed a multivariate postprocessing method which exploits the autoregressive property of the forecast errors and further extends this model using neural networks to incorporate spatial and temporal information. 

However, the basic version of AR-EMOS by \textcite{Moeller2019} shows some drawbacks. The method is estimated by two rolling training periods, one for the AR-process and one for the weights in the convex combination, respectively. Besides of the determination of appropriate lengths for the sliding windows, the model needs to be estimated for every time point again. Furthermore, it is common in the operational practice of postprocessing models to use a static instead of a rolling training period \parencite{Hess2020}. This approach can be motivated by studies as e.g. \textcite{Lang2020} outlining that the use of longer training data often leads to a better performance, independently of potential changes in the considered NWP model or the meteorological conditions. Additionally, AR-EMOS is a plug-in based model, i.e. the location and scale parameter of the predictive distribution are estimated in two steps, which can lead to a reduced forecast performance. Last but not least, the model does not take account of a general autoregressive conditional heteroscedasticity (GARCH) behavior in the distribution. 

To tackle all these issues, we utilize a modification of the smooth EMOS (SEMOS) model by \textcite{Lang2020} as baseline model. In its original formulation, the location and log-transformed scale parameter of the Gaussian distribution are described by a seasonal varying intercept and slope linked to the ensemble mean and standard deviation, respectively. Instead of using cyclic regression splines for modeling the seasonality, we use finite Fourier series. Consequently, SEMOS can be seen as a time series model explicitly considering seasonality and trend but not any autoregressive behaviour yet. Therefore, we extend SEMOS in the AR-EMOS philosophy in three directions:  

\begin{compactenum}[(i)]
    \item Assuming an autoregressive process for the forecast errors of the mean we introduce the deseasonalized autoregressive SEMOS model (DAR-SEMOS).
    \item Since the squared forecast errors of the DAR-SEMOS model show small (G)ARCH effects we adapt the DAR-SEMOS model by splitting the variance up into a seasonal variance component and an generalized autoregressive conditional variance component, resulting in the DAR-GARCH-SEMOS model.
    \item Assuming an autoregressive process for the standardized forecast errors of the SEMOS model leads to the model class we call standardized AR-SEMOS (SAR-SEMOS). In comparison to the DAR-GARCH-SEMOS model where the autoregressive behavior in the mean and variance is considered separately, this approach elegantly models the autoregressive behavior of the forecast errors and standard deviation at the same time.
\end{compactenum}

SEMOS, DAR-SEMOS, DAR-GARCH-SEMOS and SAR-SEMOS will be fitted on a static training period, facilitating more stable parameter estimates \parencite{Lang2020}. An additional advantage of the proposed methods is the fact that the parameters are estimated simultaneously by minimizing a proper scoring rule, in contrast to the two-step approach in the original AR-EMOS model. Another convenient feature of the suggested methods is their ability to postprocess forecasts of arbitrary lead time. 

To the best of the authors knowledge, the previously mentioned time series based EMOS models have not been suggested yet or further analyzed in the context of this application. Therefore, we investigate these models in comparison to the benchmark models EMOS and AR-EMOS in a case study for postprocessing of 2\,\si{\meter} surface temperature forecasts in Germany using five different lead times. Our proposed methods are able to significantly outperform the benchmark methods EMOS and AR-EMOS at the majority of the considered 283 observation stations, independently of the lead time. On the whole, the SAR-SEMOS approach yields the most noticeable improvements for all stations and lead times. Its seamless approach of jointly modelling the time series behaviour in the mean and standard deviation makes the SAR-SEMOS model appealing for practical and possibly operational use. To make our methods available for other researchers we provide an \texttt{R}-package called \texttt{tsEMOS} \parencite{Jobst2023d}.

 
The rest of the paper is organized as follows: In Section \ref{sec: Data}, the data set for our application is described. A brief overview of the benchmark ensemble postprocessing methods including a detailed description of our suggested time series based EMOS extensions is given in Section \ref{sec: Methods}. The verification methods used for the comparison of the different models are reviewed in Section \ref{sec: Verification}. In Section \ref{sec: Results}, we discuss the results of our application. We close with a conclusion and outlook in Section \ref{sec: Conclusion and outlook}.

\section{Data}
\label{sec: Data}

In our case study we locally postprocess 2\,\si{\meter} surface temperature forecasts for the five lead times 24\,\si{\hour}, 48\,\si{\hour}, 72\,\si{\hour}, 96\,\si{\hour} and 120\,\si{\hour}. The ensemble forecasts are delivered by the \textcite{ECMWF2021}, consisting of 50 perturbed ensemble members. These forecasts are initialized at 1200 UTC on a grid with resolution 0.25$^\circ$ $\times$ 0.25$^\circ$ ($\approx$ 28\,\si{km} squared). The gridded data is bilinearly interpolated to the observation stations. The 2\,\si{\meter} surface temperature observations provided by \textcite{DWD2018} contained 499 observation stations between January 2, 2015 to December 31, 2020. After selecting the stations with maximal 5\% missing observations and maximal three successive missing values at each synoptic observation station in the considered period, we obtained 283 observation stations in total. The missing observations are imputed by exponential weighted moving averaging using the \texttt{R}-package \texttt{imputeTS} by \textcite{Moritz2017}. For lead times greater than 24\,\si{\hour} we impute those observations that are missing due to the lead time in the same way as just described. AR-EMOS uses its own imputation method for these cases \parencite{Moeller2019}. 

\begin{figure}[h!]
	\begin{center}
		\includegraphics[scale = 0.5]{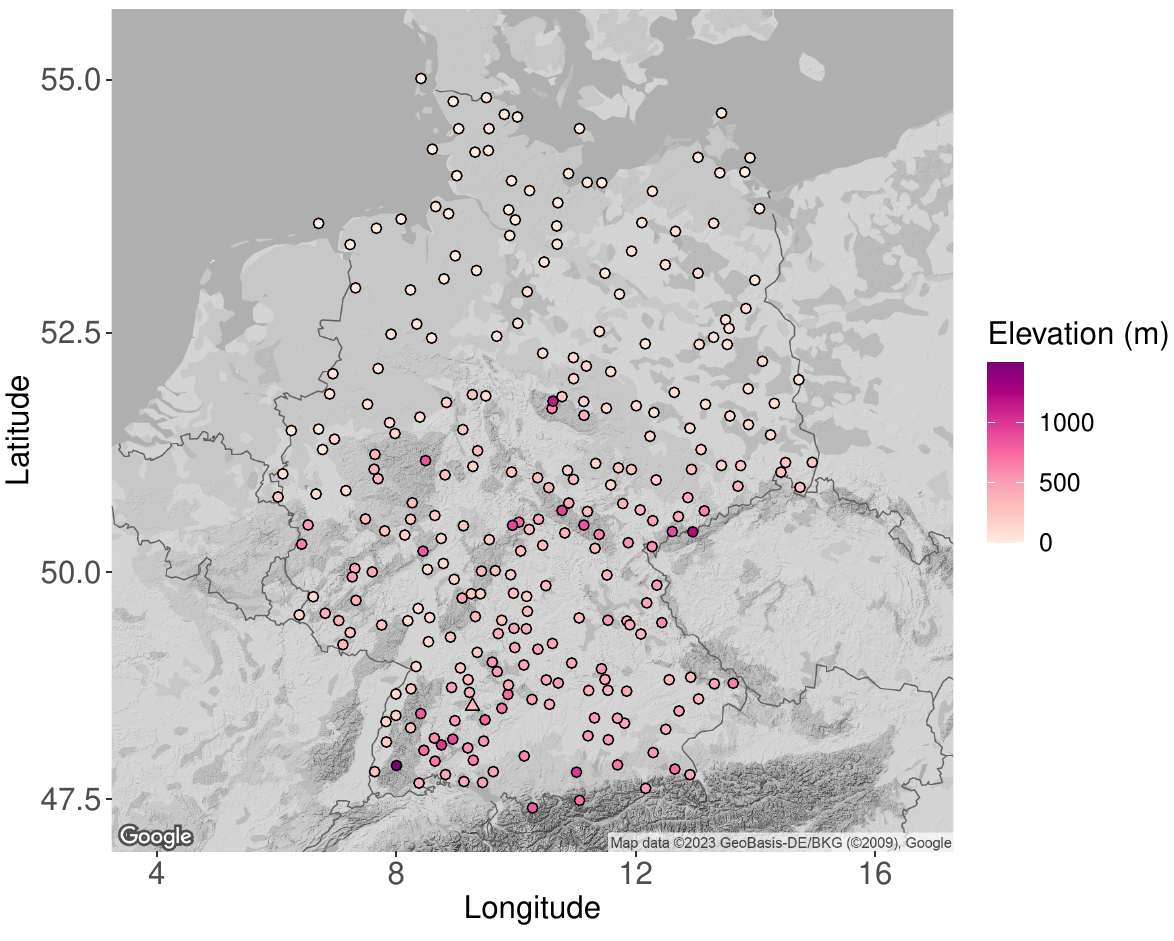}
	\end{center}	
	\caption{Observation stations for  2\,\si{\meter} surface temperature. Station Metzingen is marked by $\bigtriangleup$ in south-west Germany.}
	\label{fig: stations}
\end{figure}

In the following,
\begin{align}
\overline{x}:=\frac{1}{m}\sum\limits_{i=1}^{m}x_{i}\quad \text{and}\quad  s:=\sqrt{\frac{1}{m-1}\sum\limits_{i=1}^{m}(\overline{x}-x_{i})^2},
\end{align}
will denote the ensemble mean and standard deviation of a $m=50$ member ensemble $x_{1},\ldots, x_{m}$ for the weather quantity 2\,\si{\meter} surface temperature. The response variable 2\,\si{\meter} surface temperature is represented by $Y$ with corresponding realization $y$. 

We use the period of 2015-2019 as training set and methods driven by a sliding training window use data of 2020 as well. A comparison of all methods is based on the whole year 2020 as independent validation. The data set in use is provided by \textcite{Jobst2023e}. For the computations we apply the statistical software \texttt{R} running version 3.6.3 by \textcite{RCT2020}. More information about the implementation of the methods can be found on \texttt{\href{https://github.com/jobstdavid/paper_tsEMOS}{https://github.com/jobstdavid/paper\_tsEMOS}}.

\section{Methods}
\label{sec: Methods}

For all methods, we assume a Gaussian distribution for the weather quantity 2\,\si{\meter} surface temperature, i.e. $Y\sim \mathcal{N}(\mu, \sigma^2)$ which is reasonable and exhibits only minor differences to other distribution assumptions \parencite{Gebetsberger2019, Taillardat2021}. In the following, we shortly describe the benchmark methods EMOS, AR-EMOS using a rolling training window with training dates in 2019 and 2020. Afterwards, we explain the proposed time series extensions of the EMOS model using the static training period 2015-2019 in more detail. A summary of the model properties is given in Table \ref{tab: model properties}.

\begin{table}[h!]
\begin{center}
\resizebox{15cm}{!}{
\begin{tabular}{|c| c | c | c | c | c | c|} 
\hline
Parameter & \multicolumn{3}{c|}{$\mu$} & \multicolumn{3}{c|}{$\sigma^2$}\\ \hline
Method & season & trend & autoregression & season & trend & autoregression \\ \hline\hline
EMOS & & & & & & \\ \hline
AR-EMOS & & & $\checkmark$ & & & $\checkmark$ \\ \hline
SEMOS & $\checkmark$ & $\checkmark$ & & $\checkmark$ & $\checkmark$ & \\ \hline
DAR-SEMOS & $\checkmark$ & $\checkmark$ & $\checkmark$ & $\checkmark$ & $\checkmark$ &  \\ \hline
DAR-GARCH-SEMOS & $\checkmark$ & $\checkmark$ & $\checkmark$ & $\checkmark$ & $\checkmark$ & $\checkmark$ \\ \hline
SAR-SEMOS & $\checkmark$ & $\checkmark$ & $\checkmark$ & $\checkmark$ & $\checkmark$ & $\checkmark$\\ \hline
\end{tabular}
}
\end{center}
\caption{Overview of model specifications. If a property is directly integrated in the location $\mu$ or in the (squared) scale parameter $\sigma^2$ is indicated by a checkmark $(\checkmark)$.}
\label{tab: model properties}
\end{table}

\subsection{EMOS}
\label{sec: EMOS}

For comparison, we use the Ensemble Model Output Statistics (EMOS) proposed by \textcite{Gneiting2005}, with
\begin{align}
	\mu(t)&:=a_0+a_1\overline{x}(t),\\
	\log(\sigma(t))&:=b_0+b_1 \log(s(t)),
\end{align}
where $\overline{x}(t)$ denotes the ensemble mean, $s(t)$ the empirical ensemble standard deviation at day $t$ and coefficients $a_0,a_1,b_0,b_1\in \R$. We use a fixed rolling training period with window size of 30 days which is a common length for this model \parencites{Gneiting2005, Moeller2016} for all lead times. The coefficients of the parameters are estimated via minimization of a proper scoring rule for which we choose the continuous ranked probability score (CRPS, see Section \ref{sec: Verification}). This score yields more robust estimations as e.g. the logarithmic score according to \textcite{Gebetsberger2018}, and it can be calculated analytically.

\subsection{AR-EMOS}
\label{sec: AR-EMOS}

As further benchmark method we use the autoregressive adjusted EMOS (AR-EMOS) proposed by \textcites{Moeller2016, Moeller2019}. For each ensemble forecast $x_{i}(t)$ at day $t$, the respective error series $r_{i}(t) := Y(t) - x_{i}(t)$ is defined and an autoregressive (AR) process of order $p_i$ is fitted to each $r_i(t)$ individually. Based on the estimated parameters of the AR$(p_i)$ processes an AR-adjusted forecast ensemble is obtained via 
\begin{align}
\widetilde{x}_{i}(t):=x_i(t)+\widehat{r}_i(t)=x_i(t)+\eta_i + \sum_{j=1}^{p_{i}} \tau_{i,j}\big(\widehat{r}_i(t-j) - \eta_i\big), \label{eq: ar-emos}
\end{align}
where $\eta_i, \tau_{i,j}\in \R$, $j=1,\ldots,p_i$ are the coefficients of the respective AR$(p_i)$ process, and $\widehat{r}_i(t)$, $i=1,\ldots,m$ the corresponding residuals obtained from past observations $y(t-j)$. 
The adjusted ensemble forecasts are employed to estimate the location parameter via
\begin{align}
	\mu(t):=\frac{1}{m}\sum\limits_{i=1}^{m}\widetilde{x}_{i}(t),
\end{align}

and the scale parameter via
 
\begin{align}
	\sigma(t):=\omega\sigma_1(t)+(1-\omega)\sigma_2(t),
\end{align}
 
where $\sigma_1(t):=\sqrt{\frac{1}{m}\sum\limits_{i=1}^{m} \gamma_i^2(t)}$ is computed from the empirical variances $\gamma_i^2(t)$ of the AR$(p_i)$ process, $\sigma_2(t)$ is the empirical standard deviation of the AR-adjusted forecasts $\widetilde{x}_{i}(t)$, and $\omega\in[0,1]$ is a weight obtained by minimizing the CRPS of the predictive Gaussian distribution. We make use of the \texttt{R}-package \texttt{ensAR} proposed by \textcites{Gross2018}, where we follow their recommendation by using a fixed rolling training period of 90 days for the parameter estimation of the AR processes and additional 30 days for the weight estimation for each of the lead times.

For ensemble forecasts in $(0,24]$ \si{\hour} in advance, the residuals $\widehat{r}_i(t)$ utilized in Equation \eqref{eq: ar-emos} can directly be computed from past observations. However, for forecast horizons $24k+(0, 24]$ \si{\hour} with $k\in \N$, the residuals $\widehat{r}_i(t-k), \ldots, \widehat{r}_i(t-1)$ are not available, as $y(t-k), \ldots, y(t-1)$ have not been observed yet. To overcome this hurdle, we successively predict the missing residuals based on the recursion formula for the AR process, see \textcite{Moeller2019} for details. We will make use of this trick for the following time series models as well in those cases where the residuals are not available in practice due to the lead time.

\subsection{SEMOS}
\label{sec: SEMOS}
\newcommand\myText[1]{\text{\scriptsize\tabular[t]{@{}l@{}}#1\endtabular}}

\textcite{Lang2020} proposed a smooth EMOS model (SEMOS) for $Y\sim \mathcal{N}(\mu_S, \sigma_S^2)$ with
\begin{align}
	\mu_S(t)&:=a_0+f_0(t) + (a_1+f_1(t))\cdot \overline{x}(t)=a_0+f_0(t)+f_1(t)\cdot \overline{x}(t) + a_1\cdot \overline{x}(t),\label{eq: semos-location}\\
	\log(\sigma_S(t))&:=b_0+g_0(t) + (b_1+g_1(t))\cdot s(t)=b_0+g_0(t)+g_1(t)\cdot s(t) + b_1\cdot s(t), \label{eq: semos-scale}
\end{align}
where $f_0,f_1,g_0,g_1$ employ cyclic regression splines conditional on the day of the year $t$. We modify this approach for our situation and define
{\small
\begin{align}
	f_i(t)&:=\alpha_{i1}\sin\left(\frac{2\pi t}{365.25}\right)+\alpha_{i2}\cos\left(\frac{2\pi t}{365.25}\right)+\alpha_{i3}\sin\left(\frac{4\pi t}{365.25}\right)+\alpha_{i4}\cos\left(\frac{4\pi t}{365.25}\right),\\
	g_i(t)&:=\beta_{i1}\sin\left(\frac{2\pi t}{365.25}\right)+\beta_{i2}\cos\left(\frac{2\pi t}{365.25}\right)+\beta_{i3}\sin\left(\frac{4\pi t}{365.25}\right)+\beta_{i4}\cos\left(\frac{4\pi t}{365.25}\right),
\end{align}
}%
where $\alpha_{i,j}, \beta_{i,j}\in \R$ for $i=0,1,\, j=1,2,3,4$. To the best of our knowledge, this model has not been considered with truncated Fourier series instead of cyclic regression splines, yet. All real valued coefficients are optimized via CRPS minimization.

The selection of a truncated Fourier series to model the cyclic mean and variance behavior of surface temperature is a common approach in literature, see e.g. \textcites{Alaton2002, Campbell2005, Benth2007} for time series models and e.g. \textcites{Hemri2014, Dabernig2017, Simon2017} for the ensemble postprocessing context, where similar orders for the truncated Fourier series are selected.   

\begin{figure}[h!]
	\begin{center}
		\includegraphics[scale = 0.5]{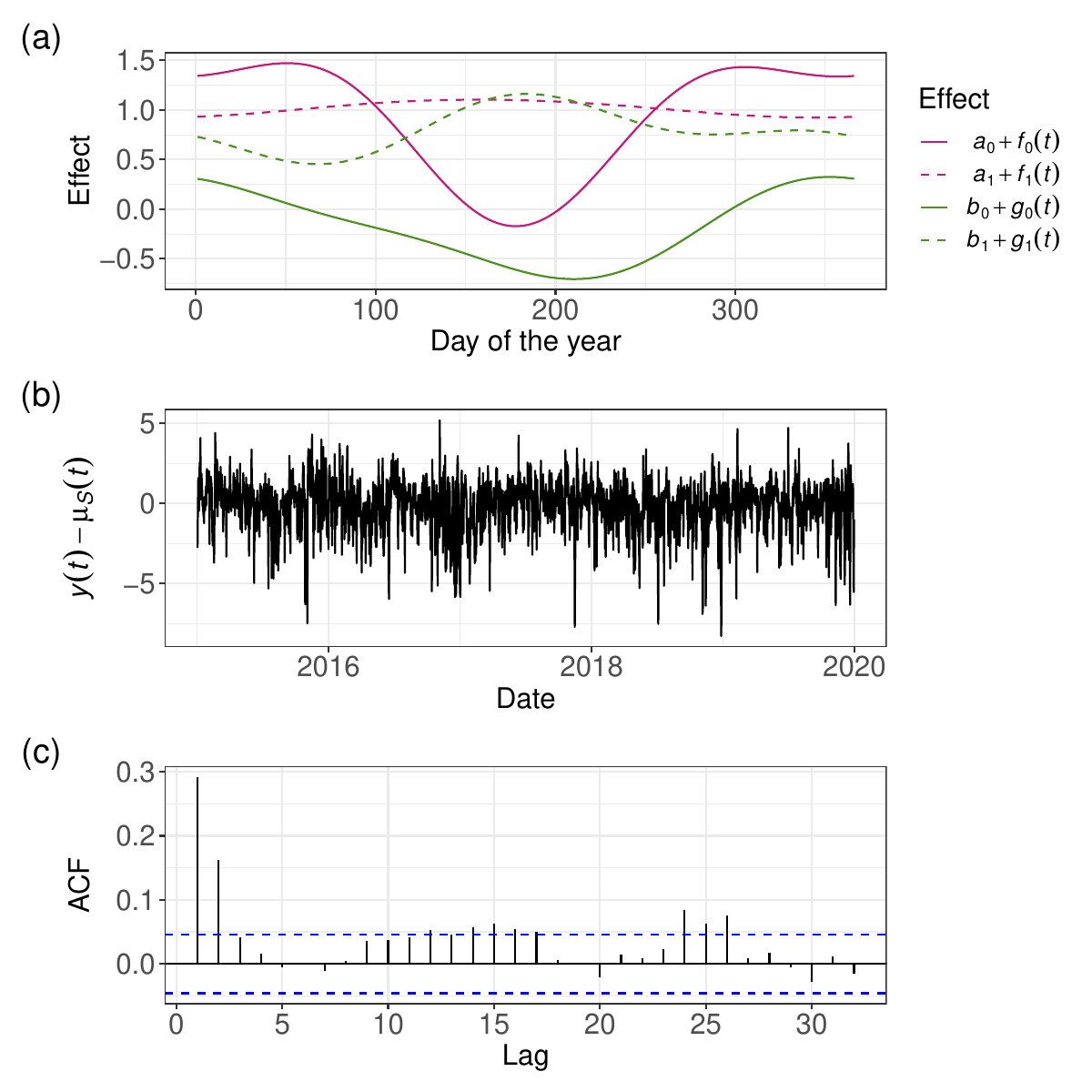}
	\end{center}	
	\caption{(a) estimated seasonal intercept and slope effects for location (purple) and log-scale (green), (b) forecast residuals, (c) autocorrelation function (ACF) for the forecast residuals at station Metzingen.}
	\label{fig: Metzingen}
\end{figure}

Considering SEMOS in a time series framework, the model explicitly accounts for seasonality in the location and $\log$-transformed scale by a seasonal effect for the intercept, $a_0+f_0(t)$ and $b_0+g_0(t)$, and a seasonal interaction effect, $f_1(t)\cdot \overline{x}(t)$ and $g_1(t)\cdot s(t)$. This becomes visible in Figure \ref{fig: Metzingen} (a) for station Metzingen, where we observe that the seasonal intercept clearly varies over the year for both distribution parameters, while the seasonal interaction effect is not that strongly pronounced in this example. By adding the terms $a_1\cdot \overline{x}(t)$ and $b_1\cdot s(t)$ to the linear predictor, which both capture the respective trend behavior, the model implicitly accounts for trend. Autoregressive effects are not accounted for, which are, however, clearly visible in Figure \ref{fig: Metzingen} (c). This motivates our extensions proposed in the following.

\subsection{DAR-SEMOS}
\label{sec: DAR-SEMOS}

\textcite{Cao2004} stated, that besides of e.g. trend behavior and seasonal cyclic patterns of the daily temperature, one should include the autoregressive property of temperature change in the daily temperature prediction, i.e. a warm day is most likely followed by another warm day and vice versa. Therefore, \textcite{Cao2004} applied an AR$(p)$ model on the daily temperature errors after removing mean and trend. In the deseasonalized autoregressive SEMOS model (DAR-SEMOS), we make use of the idea of an autoregressive error series by \textcite{Cao2004} and transfer it to the ensemble postprocessing context as in the AR-EMOS model of \textcites{Moeller2016, Moeller2019}. 

In a nutshell, DAR-SEMOS is based on the SEMOS model, where we assume an autoregressive process for the deseasonalized error series $r(t):=Y(t)-\mu_S(t)$. To be more precise, we suppose 
\begin{align}
 r(t)&:=\eta + \sum\limits_{j=1}^{p}\tau_j\big(r(t-j)-\eta\big)+\eps(t),\quad \eps(t):=\sigma(t)\cdot z(t),\label{eq: darsemos1}
\end{align}
which denotes the AR($p$) process with coefficients $\eta, \tau_j\in \R$ for $j=1,\ldots, p$. Furthermore, the AR-model error term $\eps(t)$ is specified by multiplication of $\sigma(t):=\sigma_S(t)$ as introduced in Equation \eqref{eq: semos-scale} with a white noise process $z(t) \sim \mathcal{N}(0,1)$. The parameters for the predictive distribution $\mathcal{N}(\mu, \sigma^2)$ can be recovered by 
\begin{align}
\mu(t):=\mu_S(t)+\eta + \sum\limits_{j=1}^{p}\tau_j\big(\widehat{r}(t-j)-\eta\big),\quad \sigma(t):=\sigma_S(t),
\end{align}
based on residuals $\widehat{r}(t)$ obtained from past observations. The model estimation consists of two stages, i.e. initialization and optimization of all necessary coefficients. We found that the coefficient initialization is crucial for this and all following methods in order to ensure convergence in the estimation procedure.\\

\begin{compactenum}[]
    \item \textbf{I. Initialization:} 
    \begin{enumerate}[\hspace*{0.5cm}1.]
        \item Estimate the initial coefficients for $\mu_S(t)$ by the linear regression model\\ $y(t)=\mu_S(t)+r^\ast(t)$ with white noise $r^\ast(t)\sim \mathcal{N}(0, \sigma_{r^\ast}^2)$ using the ordinary least squares method.
    \item Use the residuals $\widehat{r}^\ast(t)$ of the linear regression model to estimate the initial order $p$ and the related coefficients of the AR$(p)$ process \\ $r(t)=\eta+\sum\limits_{j=1}^{p}\tau_j\big(r(t-j)-\eta\big) + \eps^{\ast}(t)$ with white noise $\eps^{\ast}(t)\sim\mathcal{N}(0,\sigma_{\eps^{\ast}}^2)$ by the \texttt{R}-function \texttt{ar} in default settings.
        \item Set the initial coefficients for the scale parameter $\sigma_S(t)$ all to 0, except for $b_1=1$.
    \end{enumerate}
    \item \textbf{II. Optimization}:
Fix order $p$ of the AR($p$) process and optimize all coefficients simultaneously with respect to CRPS using the \texttt{R}-function \texttt{optim} with method Broyden-Fletcher-Goldfarb-Shanno (BFGS). Therefore, each optimization iteration consists of the following steps: 
    \begin{enumerate}[\hspace*{0.5cm}1.]
        \item Calculate $\widehat{\mu}_S(t)=\widehat{a}_0+\widehat{f}_0(t) +(\widehat{a}_1+\widehat{f}_1(t))\cdot \overline{x}(t)$,\quad $\widehat{r}(t)=y(t)-\widehat{\mu}_S(t)$.
        \item Predict the model residuals $\widehat{r}(t)=\widehat{\eta}+\sum\limits_{j=1}^{p}\widehat{\tau}_j\big(\widehat{r}(t-j)-\widehat{\eta}\big)$.
        \item Update $\widehat{\mu}(t)=\widehat{\mu}_S(t)+\widehat{r}(t)$,\qquad $\widehat{\sigma}(t)=\exp\left(\widehat{b}_0+\widehat{g}_0(t) + (\widehat{b}_1+\widehat{g}_1(t))\cdot s(t)\right)$.
        \item Calculate CRPS$(\mathcal{N}(\widehat{\mu}(t), \widehat{\sigma}^2(t)))$.
    \end{enumerate}
\end{compactenum}

This estimation procedure provides in contrast to the one of AR-EMOS a joint optimization of the coefficients of both parameters.

\subsection{DAR-GARCH-SEMOS}
\label{sec: DAR-GARCH-SEMOS}
Having a look at the  squared residuals $\widehat{\eps}^2(t)$ of the DAR-SEMOS models in Table \ref{tab: significance residuals0}, we can observe (G)ARCH effects. This is in line with the results of \textcites{Campbell2005, Benth2012a}, both detecting these effects for temperature time series, as well. Therefore, one might include this behavior in the DAR-SEMOS model, too. Consequently, we extended the DAR-SEMOS model by a multiplicative and an additive version for the variance, as proposed by \textcites{Benth2012a, Campbell2005}, respectively. Due to minor differences among both model formulations, we restrict ourselves to the multiplicative version, called DAR-GARCH-SEMOS, in the following.

\begin{table}[h!]
\begin{center}
\begin{tabular}{c c  c  c c c c } 
\toprule
Lead time & $k=1$ & $k=2$ & $k=3$ & $k=4$ & $k=5$ & $k=10$ \\ \hline
24\,\si{\hour} &  80.57 & 78.80 & 77.03 & 77.74 & 75.62 & 73.50 \\ 
48\,\si{\hour} &  81.98 & 83.04 & 80.21 & 79.86 & 78.80 & 73.14 \\ 
72\,\si{\hour} &  81.27 & 80.57 & 78.09 & 75.62 & 75.27 & 69.61 \\ 
96\,\si{\hour} &  84.81 & 82.33 & 74.20 & 78.09 & 80.21 & 75.62 \\ 
120\,\si{\hour} &  76.68 & 76.33 & 69.26 & 82.33 & 84.10 & 83.04 \\ 
\bottomrule
\end{tabular}
\end{center}
\caption{Significance of dependence (in \%) over all stations in the  squared residuals $\widehat{\eps}^2(t)=(y(t)-\widehat{\mu}(t))^2$ of the estimated DAR-SEMOS models up to lag $k$. The $p$-values of the performed Ljung-Box test were corrected by the Benjamini-Hochberg procedure with significance level of $\alpha=0.05$.}
\label{tab: significance residuals0}
\end{table}

The DAR-GARCH-SEMOS has basically the same model assumptions, as DAR-SEMOS in Equation \eqref{eq: darsemos1} only with a slightly different parametrization of the variance $\sigma^2(t)$ to account for seasonality and GARCH effects. Therefore, the variance $\sigma^2(t)$  consists of a seasonal variance factor $\sigma_S^2(t)$ and a conditional variance factor $\sigma_G^2(t)$, i.e.
\begin{align}
	\sigma^2(t):=\sigma_S^2(t)\cdot \sigma_G^2(t),
\label{eq: DAR-GARCH-mult}	
\end{align}
where $\sigma_S(t)$ is specified as in Equation \eqref{eq: semos-scale}. Inserting $\sigma(t)$ of Equation \eqref{eq: DAR-GARCH-mult} into the error series $\eps(t)=\sigma(t)\cdot z(t)$ of Equation \eqref{eq: darsemos1} yields
\begin{align}
	\eps(t)=\sigma_S(t)\cdot \sigma_G(t)\cdot z(t)\quad \Leftrightarrow\quad \underbrace{\frac{\eps(t)}{\sigma_S(t)}}_{=:\rho(t)}=\sigma_G(t)\cdot z(t).
\end{align}
Due to simplicity, we assume $\rho \sim \text{GARCH}(1,1)$ with
\begin{align}
	\sigma_G^2(t):=\omega_0+\omega_1\sigma_G^2(t-1)+\omega_2\rho^2(t-1), \label{eq: GARCH}
\end{align}
where $\omega_0, \omega_1, \omega_2\geq 0$. Different orders for $r,s$ of the GARCH$(r,s)$ model are left for further research. The parameters for the predictive distribution $\mathcal{N}(\mu, \sigma^2)$ can be derived by 
\begin{align}
\mu(t):=\mu_S(t)+\eta + \sum\limits_{j=1}^{p}\tau_j\big(\widehat{r}(t-j)-\eta\big),\quad \sigma(t):=\sigma_S(t)\cdot \sigma_G(t),
\end{align}
with residuals $\widehat{r}(t)$ obtained from past observations. Analogously to DAR-SEMOS, the estimation of the DAR-GARCH SEMOS model consists of an initialization and an optimization step:

\begin{compactenum}[]
    \item \textbf{I. Initialization:} 
    \begin{enumerate}[\hspace*{0.5cm}1.]
        \item Perform the initialization steps 1. and 2. as in DAR-SEMOS to obtain $\mu_S(t)$ and the initial order as well as coefficients of the $\mathrm{AR}(p)$ process. Additionally, calculate the residuals $\widehat{\eps}^{\ast}(t)$ of the initial AR($p$) process.
        \item Estimate the initial coefficients for $\sigma_S(t)$ by the linear regression model\\ $\log(\widehat{s}(t))=\log(\sigma_S(t))+\tilde{r}(t)$ with empirically observed standard deviation $\widehat{s}(t)$ and white noise $\tilde{r}(t)\sim \mathcal{N}(0, \sigma_{\tilde{r}}^2)$ using the ordinary least squares method. The quantity $\widehat{s}(t)$ is obtained by calculating the empirical standard deviation based on the observations of symmetric training windows around the day of the year $t$.
        \item Calculate the deseasonalized residuals $\displaystyle \widehat{\rho}(t)=\frac{\widehat{\eps}^{\ast}(t)}{\widehat{\sigma}_S(t)}$ with initial residuals $\widehat{\eps}^{\ast}(t)$ for which a GARCH$(1,1)$ model is fitted to get the corresponding initial values $\sqrt{\widehat{\omega}}_0, \sqrt{\widehat{\omega}}_1, \sqrt{\widehat{\omega}}_2$ using the \texttt{R}-package \texttt{rugarch} by \textcite{Ghalanos2022}.
    \end{enumerate}
    \item \textbf{II. Optimization}:
Fix order $p$ of the AR($p$) process and optimize all coefficients of both parameters with the same settings as for DAR-SEMOS: 
    \begin{enumerate}[\hspace*{0.5cm}1.]
        \item Perform optimization steps 1. and 2. as in DAR-SEMOS to obtain $\widehat{\mu}_S(t)$ and $\widehat{r}(t)$.
        \item Calculate the  residuals $\displaystyle \widehat{\rho}(t)=\frac{\widehat{\eps}(t)}{\widehat{\sigma}_S(t)}$ with updated $$\widehat{\sigma}_S(t)=\exp\left(\widehat{b}_0+\widehat{g}_0(t) + (\widehat{b}_1+\widehat{g}_1(t))\cdot s(t)\right),\, \widehat{\eps}(t)=\widehat{r}(t)-\widehat{\eta}-\sum\limits_{j=1}^{p}\widehat{\tau}_j\big(\widehat{r}(t-j)-\widehat{\eta}\big).$$
          \item Update $\widehat{\sigma}_G^2(t)=\widehat{\omega}_0^2+\widehat{\omega}_1^2\widehat{\sigma}_G^2(t-1)+\widehat{\omega}_2^2\widehat{\rho}^2(t-1)$, with squared coefficients $\widehat{\omega}_0, \widehat{\omega}_1, \widehat{\omega}_2$, to ensure their non-negativity.
        \item Update $\widehat{\mu}(t)=\widehat{\mu}_S(t)+\widehat{r}(t)$,\qquad $\widehat{\sigma}(t)=\widehat{\sigma}_S(t)\cdot \widehat{\sigma}_G(t)$.
	   \item Calculate CRPS$(\mathcal{N}(\widehat{\mu}(t), \widehat{\sigma}^2(t)))$.
    \end{enumerate}
\end{compactenum}

\subsection{SAR-SEMOS}
\label{sec: SAR-SEMOS}
All previously mentioned methods consider the autoregressive behavior of the forecast errors and the variance separately. However, modeling the autoregressive behavior of the standardized errors 
\begin{align}
z(t):=\frac{Y(t)-\mu(t)}{\sigma(t)},\label{eq:sar-semos1}
\end{align}
helps to overcome this potential issue and was first considered by \textcite{Wakaura2007}, and later again by \textcite{Schiller2012} for temperature models. Therefore, we follow this approach for the standardized AR-SEMOS (SAR-SEMOS), specify $\mu(t):=\mu_S(t)$ and $\sigma(t):=\sigma_S(t)$ as in Equations \eqref{eq: semos-location}, \eqref{eq: semos-scale} and extend the SEMOS model via

\begin{align}
z(t):=\eta + \sum\limits_{j=1}^{p}\tau_j\big(z(t-j)-\eta\big)+\xi(t)\label{eq:sar-semos2},
\end{align}

where $z\sim \text{AR}(p)$ with coefficients $\eta, \tau_j\in \R$ for $j=1,\ldots, p$, and $\xi(t)\sim \mathcal{N}(0,\sigma_\xi^2)$ is white noise. The parameters for the predictive distribution $\mathcal{N}(\mu, \sigma^2)$ can be obtained by 
\begin{align}
\mu(t):=\mu_S(t)+\sigma_S(t)\cdot \left(\eta +\sum\limits_{j=1}^{p}\tau_j\big(\widehat{z}(t-j)-\eta\big)\right),\quad \sigma(t):=\sigma_S(t),
\end{align}
with residuals $\widehat{z}(t)$ obtained from past observations. Analogously to DAR-SEMOS, the estimation of the SAR-SEMOS model consists of an initialization and an optimization step.

\begin{compactenum}[]
    \item \textbf{I. Initialization:} 
    \begin{enumerate}[\hspace*{0.5cm}1.]
    \item Perform the initialization steps 1. of DAR-SEMOS and step 2. of DAR-GARCH-SEMOS to obtain $\widehat{\mu}_S(t)$ and $\widehat{\sigma}_S(t)$.
    \item Use the standardized residuals $\widehat{z}(t)=\frac{y(t)-\widehat{\mu}_S(t)}{\widehat{\sigma}_S(t)}$ to estimate the initial order $p$ and the related coefficients of the residual AR$(p)$ process in Equation \eqref{eq:sar-semos2} by the \texttt{R}-function \texttt{ar} in default settings.
    \end{enumerate}
    \item \textbf{II. Optimization}:
Fix order $p$ of the AR($p$) process and optimize all coefficients of both parameters in Equations \eqref{eq:sar-semos1}, \eqref{eq:sar-semos2} simultaneously: 
\begin{enumerate}[\hspace*{0.5cm}1.]
 \item Update $$\widehat{\mu}_S(t)=\widehat{a}_0+\widehat{f}_0(t) + (\widehat{a}_1+\widehat{f}_1(t))\cdot \overline{x}(t),\, \widehat{\sigma}_S(t)=\exp\left(\widehat{b}_0+\widehat{g}_0(t) + (\widehat{b}_1+\widehat{g}_1(t))\cdot s(t)\right).$$
        \item Update $\widehat{z}(t)=\frac{y(t)-\widehat{\mu}_S(t)}{\widehat{\sigma}_S(t)}$ and predict the model residuals $$\widehat{z}(t)=\widehat{\eta}+\sum\limits_{j=1}^{p}\widehat{\tau}_j\big(\widehat{z}(t-j)-\widehat{\eta}\big).$$
        \item Update $\widehat{\mu}(t)=\widehat{\mu}_S(t)+ \widehat{\sigma}_S(t)\cdot \widehat{z}(t)$, $\widehat{\sigma}(t):=\widehat{\sigma}_S(t)$.
        \item Calculate CRPS$(\mathcal{N}(\widehat{\mu}(t), \widehat{\sigma}^2(t)))$.
    \end{enumerate}
\end{compactenum}

\section{Verification}
\label{sec: Verification}

\paragraph{Assessing calibration and sharpness.}
As argued by \textcite{Gneiting2005, Gneiting2007}, the general aim of probabilistic forecasting is to maximize sharpness of the predictive distribution subject to calibration. \textit{Calibration} refers to the statistical consistency between the predictive cumulative distribution function (CDF) $F$ and the associated observation $Y$. Consequently it is a joint property of the forecasts and the verifications. \textit{Sharpness} refers to the concentration of the predictive distribution $F$ and is a property of the probabilistic forecasts only. The more concentrated the forecast, the sharper the forecast, and the sharper the better, subject to calibration. In the following, we will present methods to measure calibration and sharpness which will be used in the subsequent application. 

\begin{figure}[h!]
	\begin{center}
		\includegraphics[scale = 0.6]{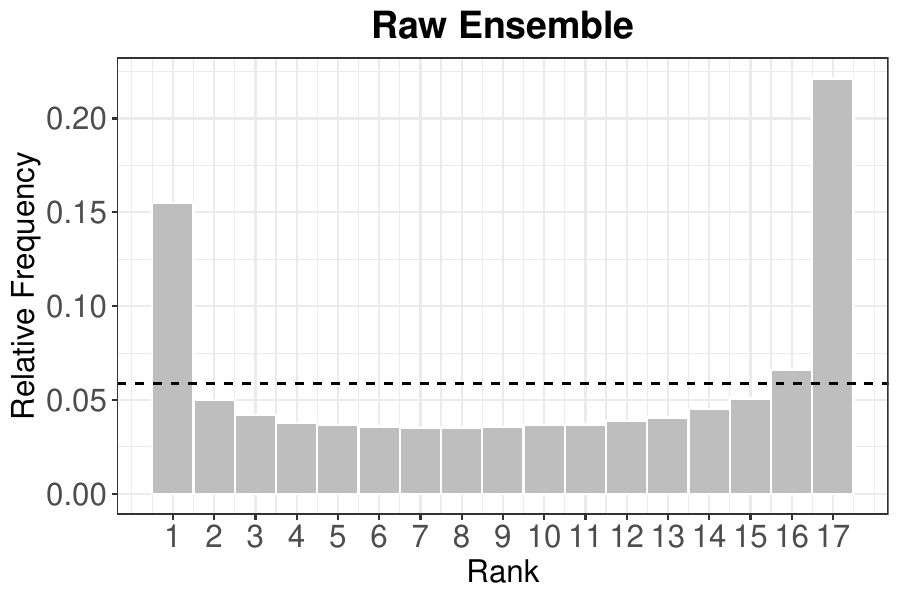}
	\end{center}	
	\caption{Verification rank histogram of the raw ensemble aggregated over all time points, stations and lead times in the validation period.}
	\label{fig: vr_hist}
\end{figure}

A continuous predictive probabilistic forecast $F$ is calibrated if $F(Y)$ is uniformly distributed \parencites{Dawid1984, Gneiting2007a}. For the assessment of the calibration a so called probability integral transform (PIT) histogram can be used as visual tool, where the PIT values are obtained by evaluating the predictive CDF $F$ at the observations. Any departures from uniformity of the PIT histogram can indicate that the predictive distribution $F$ is miscalibrated in some way. If the predictive distribution $F$ is calibrated, the PIT values should have variance $\mathrm{Var}(U)=1/12\approx 0.0833$, representing neutral dispersion. Values smaller/greater than 0.0833 indicate overdispersion/underdispersion of the predictive distribution $F$ according to \textcite{Gneiting2013}. 

A discrete counterpart of the PIT histogram is the so called verification rank histogram displaying the histogram of ranks of observations with respect to the corresponding ordered ensemble forecasts \parencite{Talagrand1997}. For a calibrated $m$-member ensemble, the ranks should be uniformly distributed on the set $\{1,\ldots, m+1\}$.

\paragraph{Central prediction interval.}
Calibration of a predictive distribution can also be assessed by the (empirical) coverage of a $(1-\alpha)\cdot 100\%$ central prediction interval $[l,u]$ for $\alpha\in (0,1)$, where $l$ and $u$ denote the quantiles of the predictive distribution $F$ at quantile level $\frac{\alpha}{2}$ and $1-\frac{\alpha}{2}$, respectively  \parencite{Gneiting2007}. The coverage of a $(1-\alpha)\cdot 100\%$ central prediction interval is the proportion of validating observations between the lower and upper $\frac{\alpha}{2}$-quantiles of the predictive distribution. If a predictive distribution is calibrated, then $(1-\alpha)\cdot 100\%$ of observations should fall within the range of the central prediction interval.
Sharpness of a predictive distribution can be validated by the width of a $(1-\alpha)\cdot 100\%$ central prediction interval \parencite{Gneiting2007}. Sharper distributions correspond to narrower prediction intervals. In the case of a $m$-member forecast ensemble, we consider an $\frac{m-1}{m+1}\cdot 100\%$ central prediction interval which corresponds to the nominal coverage of the raw forecast ensemble, thus allowing a direct comparison of all probabilistic forecasts. The target coverage rate for a $m=50$ member ensemble is approximately 96.08\%.

\paragraph{Proper scoring rules.}
Above, we have introduced tools to analyze  calibration and sharpness of a probabilistic forecast separately. \textit{Proper scoring rules} assess calibration and sharpness properties simultaneously and play therefore important roles in the comparative evaluation and ranking of competing forecasts \parencite{Gneiting2007a}.
A proper scoring rule which we use in the following case study, is the \textit{continuous ranked probability score} (CRPS) \parencite{Matheson1976}. It is defined as 
\begin{align}
\text{CRPS}(F,y):=\int\limits_{-\infty}^{\infty}(F(z)-\mathds{1}\{z\geq y\})^2\, \mathrm{d}z,	\label{eq: CRPS1}
\end{align}
where $F$ is the predictive CDF, $y$ is the true/observed value and $\mathds{1}$ denotes the indicator function. For the case of $F\sim\mathcal{N}(\mu, \sigma^2)$, \textcite{Gneiting2005} derived the closed form 
\begin{align}
    \text{CRPS}(F,y)=\sigma\left(\frac{y-\mu}{\sigma}\left(2\Phi\left(\frac{y-\mu}{\sigma}\right)-1\right)+2\varphi\left(\frac{y-\mu}{\sigma}\right)-\frac{1}{\sqrt{\pi}}\right),
\end{align}
where $\varphi$ and $\Phi$ denote the PDF and CDF of the standard normal distribution, respectively. The mean CRPS over a set of forecast cases will be denoted by $\overline{\text{CRPS}}$.

As additional proper scoring rule, we consider the \textit{logarithmic score} (LogS), which is defined as 
\begin{align}
\text{LogS}(F,y):=-\log(f(y)),
\end{align}
where $f$ denotes the density of the predictive distribution $F$.

\paragraph{Consistent scoring functions.}
In practice, a probabilistic forecast is sometimes reduced to a point forecast via a statistical summary function such as the mean or median. In this situation, \textit{consistent scoring functions} provide useful tools for forecast evaluation and generate proper scoring rules \parencite{Gneiting2011b}. In particular, we use the \textit{squared error} 
\begin{align}
\text{SE}(F,y):=(\text{mean}(F)-y)^2,	
\end{align}
for the mean forecast mean$(F)$. For $n$ forecast cases we get the \textit{root mean squared error} via 
\begin{align}
\text{RMSE}:=\sqrt{\frac{1}{n}\sum\limits_{i=1}^{n}\text{SE}(F_i,y_i)}.
\end{align}

\paragraph{Relative improvement.}
To assess the relative improvement of a forecast with respect to a given reference forecast, one can calculate the \textit{continuous ranked probability skill score} (CRPSS) via
\begin{align}
	\text{CRPSS}:=1-\frac{\overline{\text{CRPS}}}{\overline{\text{CRPS}}_{\text{ref}}},
\end{align}
where $\overline{\text{CRPS}}_\text{ref}$ denotes the $\overline{\text{CRPS}}$ of the reference forecast.

\paragraph{Statistical tests.}
To evaluate the statistical significance of the differences in performance between two competing postprocessing methods, we apply a \textit{Diebold-Mariano test} \parencite{Diebold1995} to the verification score time series of both methods separately for each station and lead time case. Afterwards we use the \textit{Benjamini-Hochberg procedure} \parencite{Benjamini1995} suggested by \textcite{Wilks2016}, which allows to account for multiple testing regarding different stations, and controls the overall probability of a type I error, for which we choose $\alpha = 0.05$ in the subsequent analysis. 

Furthermore, to test for possible autocorrelations in the (squared) forecast residuals, we apply the Ljung-Box test by \textcite{Ljung1978}, which allows to test for the absence of serial autocorrelation, up to a specified lag $k$. We use the standard \texttt{R}-function \texttt{Box.test} to make these checks. In case of multiple testing, we perform again the Benjamini-Hochberg procedure to control the overall probability of a type I error with $\alpha = 0.05$.

The verification of the methods is carried out using the \texttt{R}-package \texttt{eppverification} by \textcite{Jobst2021}.

\section{Results}
\label{sec: Results}

In the following subsections, we first evaluate the results of all models over all stations and lead times with respect to the scores and methods mentioned in Section \ref{sec: Verification}. Furthermore, we investigate the order $p$ of the AR-process and we examine the remaining autocorrelation in the (squared) residuals. Afterwards, we discuss the lead time-specific results of all models. In the last subsection, we outline the station-specific results as well as statistical significance of our suggested models in comparison to EMOS, AR-EMOS and SEMOS. 

\subsection{General results}
\label{sec: General results}

\begin{figure}[h!]
	\begin{center}
		\includegraphics[scale = 0.6]{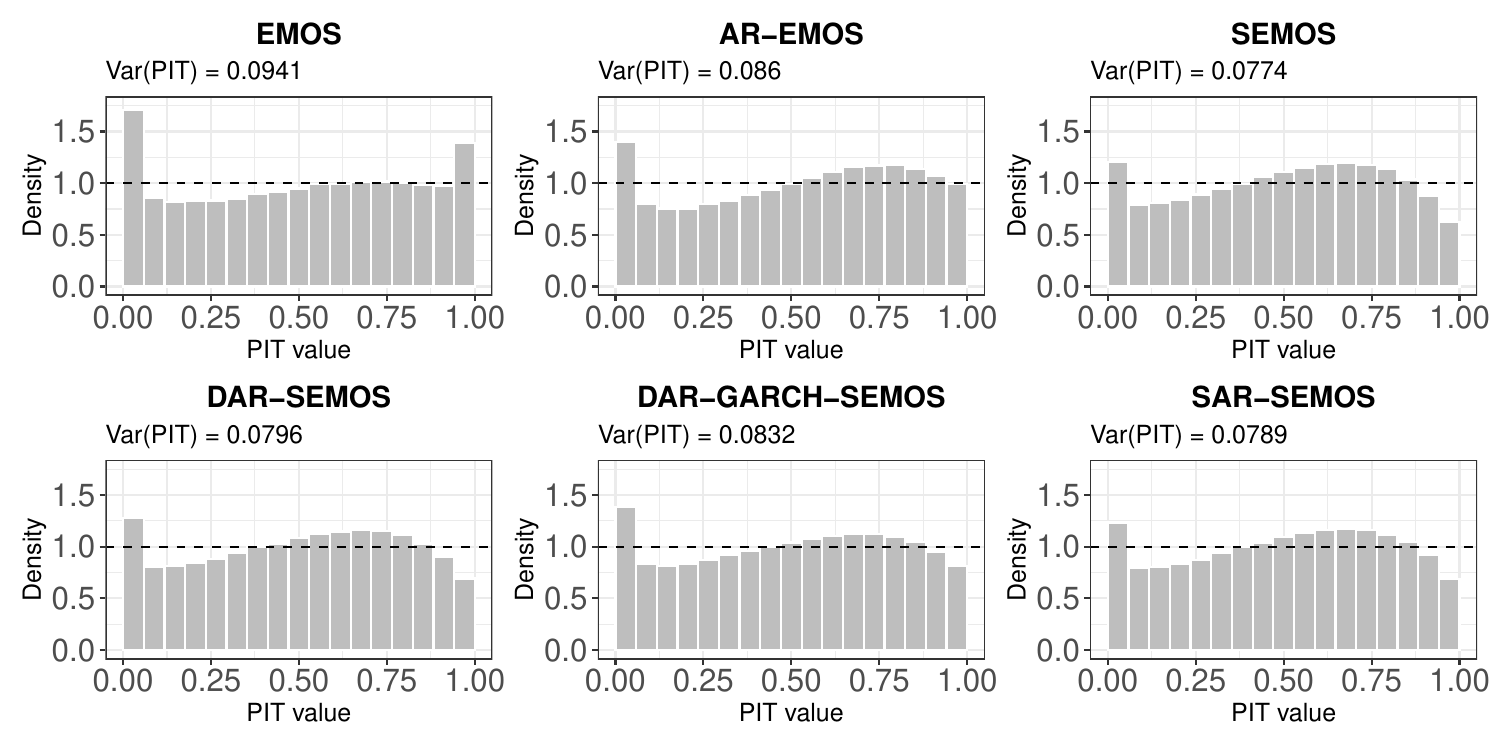}
	\end{center}	
	\caption{PIT histograms of the methods aggregated over all time points, stations and lead times in the validation period.}
	\label{fig: pit_hist}
\end{figure}

Figure \ref{fig: pit_hist} shows the PIT histograms of the models considered in Section \ref{sec: Methods}. All methods clearly improve the calibration in comparison to the raw ensemble, which shows a strong underdispersion in the verification rank histogram in Figure \ref{fig: vr_hist}. The PIT histograms of all methods exhibit a slight underdispersion at the left end, indicating that the lower tail of the distribution is a bit too light. Furthermore, all PIT histograms, except of the one for EMOS, exhibit a small bump shape approximately at PIT value 0.75, which outlines a slight overdispersion. Looking at the variance of the PIT values, DAR-GARCH-SEMOS yields a value of 0.082 which is the closest to the reference variance of 0.0833 indicating neutral dispersion, where the variance of the PIT values of the other methods slightly fluctuate around this value. With respect to the coverage assessing calibration, SEMOS yields the closest value to the nominal coverage of 95.25\%.

\begin{table}[h!]
\begin{center}
\begin{tabular}{c c  c c c c} 
\toprule
Method & CRPS & LogS & RMSE & Width & Coverage\\ \hline 
Raw ensemble & 1.165 & $-$ & 2.060 & 4.935 & 74.69\\ \hline
EMOS & 1.007 & 2.183 & 1.899 & 6.407 & 90.08\\
AR-EMOS & 0.943 & 1.952 & 1.787 & 6.442 & 93.41\\
SEMOS & 0.908 & 1.880 & 1.747 & 6.688 & \textbf{95.25}\\ \hline
DAR-SEMOS & 0.902 & 1.885 & 1.734 & 6.458 & 94.67\\
DAR-GARCH-SEMOS & 0.902 & 1.906 & 1.735 & \textbf{6.282} & 93.93\\
SAR-SEMOS & \textbf{0.890} & \textbf{1.865} & \textbf{1.711} & 6.401 & 94.94\\
\bottomrule
\end{tabular}
\end{center}
\caption{Verification scores aggregated over all time points, stations and lead times in the validation period. Bold value represents the best value for each score.}
\label{tab: total scores}
\end{table}

Further looking at Table \ref{tab: total scores}, we observe that all methods improve the raw ensemble with respect to CRPS by around 14\%-24\% and RMSE by around 8\%-17\%. In terms of the previously mentioned scores including the LogS, SEMOS is already able to obviously outperform EMOS and AR-EMOS. This shows, that a proper modeling of the seasonality for methods using longer training data can yield to tremendous enhancements in comparison to methods using rolling training periods. Furthermore, the proposed models DAR-SEMOS, DAR-GARCH-SEMOS and SAR-SEMOS show further improvements over SEMOS, while SAR-SEMOS yields the lowest CRPS, LogS and RMSE of all considered methods. Last but not least DAR-GARCH-SEMOS provides the sharpest forecasts indicated by the lowest width, while this comes along with a loss in calibration in comparison to new proposed time series based models. 

The results underline, that the inclusion of an autoregressive process for the forecast errors and an autoregressive conditional variance can improve the forecasting performance even more. Similar to \textcite{Moeller2016}, we found that the lower orders $p=1,2,3$ for the AR($p$) process are mostly chosen and sufficient for DAR-SEMOS, DAR-GARCH-SEMOS and SAR-SEMOS. Furthermore, AR(2) is the most prominent type of AR-structure, followed by AR(1) and AR(3). 
Table \ref{tab: significance residuals1} highlights for these three extensions, that they effectively remove the autocorrelation in the residuals $\widehat{\eps}(t)$. Moreover, Table \ref{tab: significance residuals2} outlines that there are (G)ARCH effects in the squared residuals $\widehat{\eps}^2(t)$ of DAR-SEMOS and SAR-SEMOS left, while DAR-GARCH-SEMOS removes nearly all of these effects. Moreover, modeling the autoregressive behavior of the forecast residuals and standard deviation simultaneously seems to be more effective than modeling the autoregression of both quantities separately with respect to the considered scores. 

\begin{table}[h!]
\begin{center}
\begin{tabular}{c c  c  c c c c c} 
\toprule
Method & $k=1$ & $k=5$ & $k=10$ & $k=15$ & $k=20$ & $k=25$ & $k=30$ \\ \hline
DAR-SEMOS & 0.00 & 0.00 & 0.00 & 0.07 & 0.07 & 0.07 & 0.21 \\ 
DAR-GARCH-SEMOS &  0.00 & 0.00 & 0.14 & 0.14 & 0.14 & 0.21 & 0.42 \\ 
SAR-SEMOS & 0.00 & 0.00 & 0.00 & 0.00 & 0.00 & 0.00 & 0.00 \\ 
\bottomrule
\end{tabular}
\end{center}
\caption{Significance of dependence (\%) over all stations and lead times in the  residuals $\widehat{\eps}(t)$ of the estimated methods up to lag $k$. The $p$-values of the performed Ljung-Box test were corrected by the Benjamini-Hochberg procedure with significance level $\alpha = 0.05$.}
\label{tab: significance residuals1}
\end{table}

\begin{table}[h!]
\begin{center}
\begin{tabular}{c c  c  c c c c } 
\toprule
Method & $k=1$ & $k=2$ & $k=3$ & $k=4$ & $k=5$ & $k=10$ \\ \hline
DAR-SEMOS & 80.99 & 80.14 & 75.69 & 78.87 & 78.80 & 74.77 \\ 
DAR-GARCH-SEMOS &  17.17 & 16.75 & 16.11 & 16.54 & 15.12 & 14.56 \\ 
SAR-SEMOS &  68.90 & 69.12 & 67.35 & 67.77 & 66.50 & 59.08 \\ 
\bottomrule
\end{tabular}
\end{center}
\caption{Significance of dependence (\%) over all stations and lead times  in the squared residuals $\widehat{\eps}^2(t)$ of the estimated methods up to lag $k$. The $p$-values of the performed Ljung-Box test were corrected by the Benjamini-Hochberg procedure with significance level $\alpha = 0.05$.}
\label{tab: significance residuals2}
\end{table}

\subsection{Lead time-specific results}
\label{sec: Lead time-specific results}

\begin{figure}[h!]
\centering
\begin{subfigure}[b]{0.8\textwidth}
   \includegraphics[width=1\linewidth]{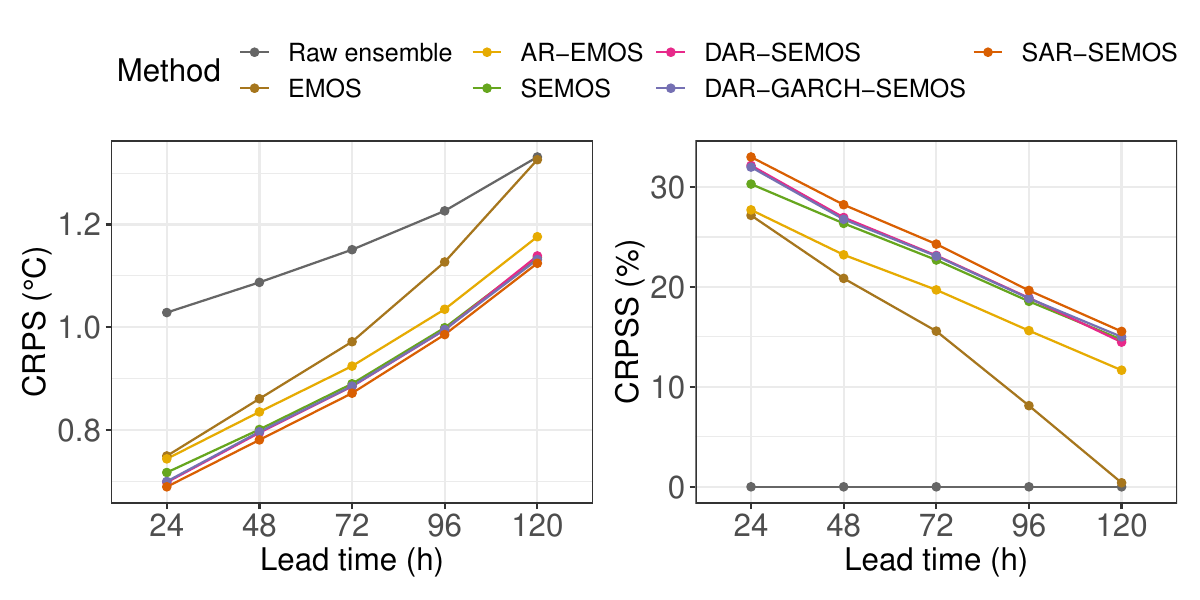}
   \caption{CRPS (left) and CRPSS over raw ensemble (right).}
\end{subfigure}

\begin{subfigure}[b]{0.8\textwidth}
   \includegraphics[width=1\linewidth]{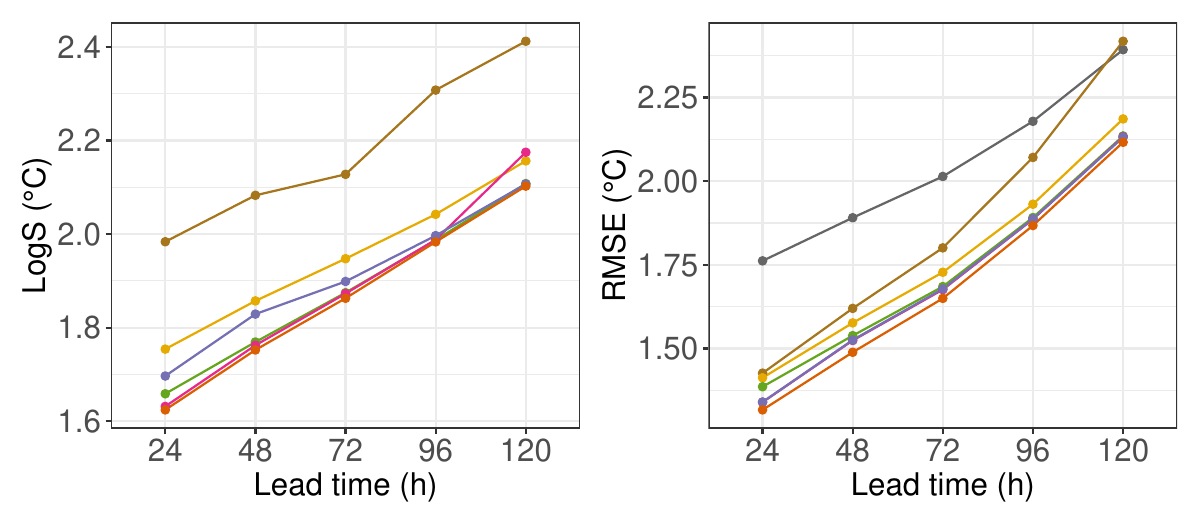}
   \caption{LogS (left) and RMSE (right).}
\end{subfigure}

\begin{subfigure}[b]{0.8\textwidth}
   \includegraphics[width=1\linewidth]{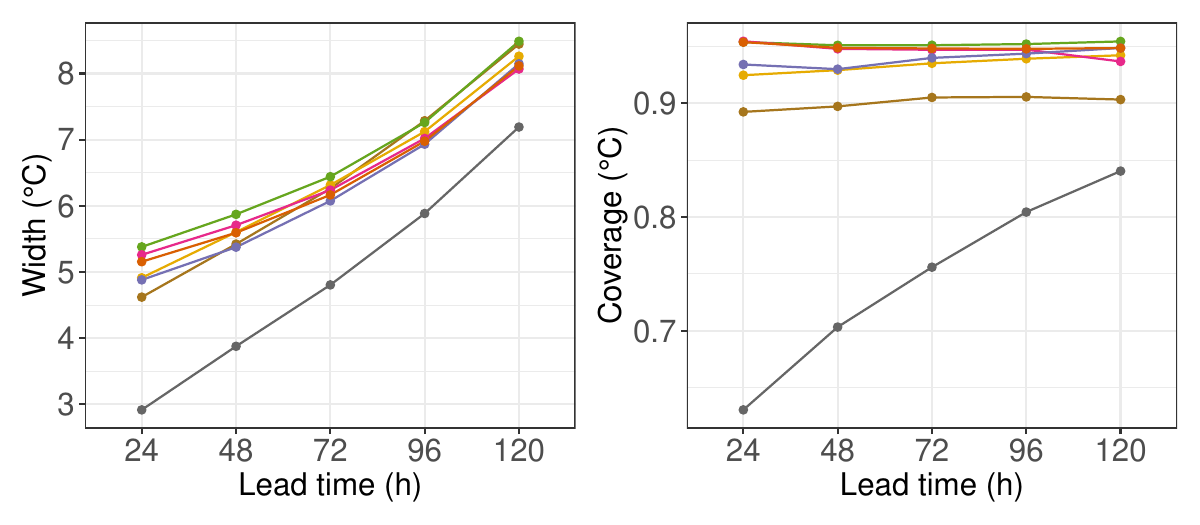}
   \caption{Width (left) and coverage (right).}
\end{subfigure}

\caption[]{Lead time specific scores for all methods aggregated over all time points and stations in the validation period.}
\label{fig: lt_plots}
\end{figure}

For the investigation of an effect of the lead time on predictive performance of the methods we consider several verification scores depending on the lead time in Figure \ref{fig: lt_plots}. In terms of CRPS, LogS and RMSE, SAR-SEMOS yields the lowest, and therefore best scores for all lead times, followed by SEMOS, DAR-SEMOS and DAR-GARCH-SEMOS. In comparison to the  benchmark methods EMOS and AR-EMOS there are pronounced differences with respect to the mentioned scores independently of the lead time, indicating that our models provide a strong improvement.

 All in all, these scores become larger with increasing lead time, as the raw ensemble yields a poorer performance. However, with increasing lead time the performance differences between SEMOS and the methods DAR-SEMOS, DAR-GARCH-SEMOS, SAR-SEMOS decrease. This effect can possibly be led back to the fact, that the higher the lead time is, the more residuals need to be predicted for the time series based EMOS extensions. Consequently, its forecasts become less accurate. Nonetheless, for example SAR-SEMOS still yields CRPSS improvements of 4\%, 3\%, 2\%, 1\%, 1\% in comparison to SEMOS for the five considered lead times, respectively, underlining its superiority over the other methods. 

Considering the width, DAR-GAR-SEMOS yields the sharpest forecasts over all lead times, while SEMOS has the highest width, introducing a higher uncertainty into the forecasts. However, this yields to more calibrated forecasts for SEMOS and less calibrated forecasts for DAR-GARCH-SEMOS. Our other extensions show a similar performance and outperform EMOS and AR-EMOS as well, almost independent of the lead time.

\subsection{Station-specific results and statistical significance}
\label{sec: Station-specific results and statistical significance}

Additionally to the effect of the lead time, we investigate the station-specific performance of DAR-SEMOS, DAR-GARCH-SEMOS and SAR-SEMOS over SEMOS, as latter method has already proven that it clearly outperforms EMOS and AR-EMOS in terms of CRPS as well as CRPSS (see Figure \ref{fig: lt_plots}).

\begin{figure}[h!]
	\begin{center}
		\includegraphics[scale = 0.7]{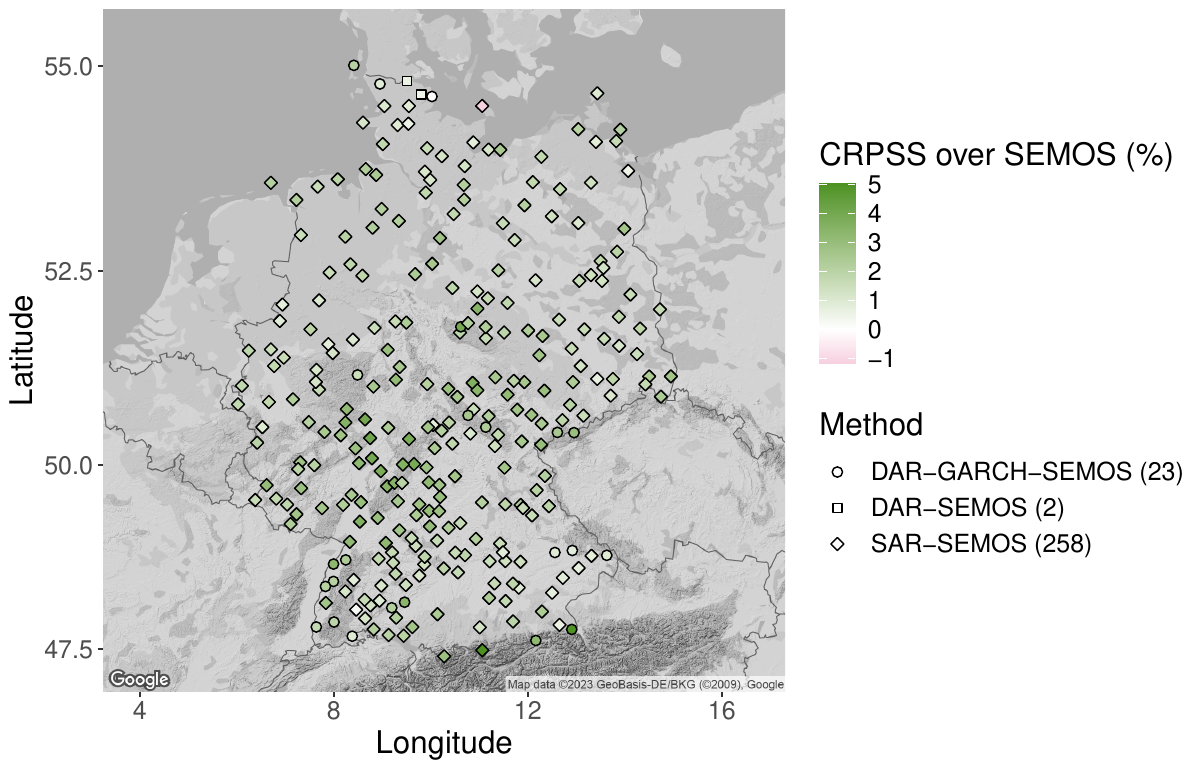}
	\end{center}	
	\caption{Station-wise highest CRPSS of the considered methods across all lead times over SEMOS in \% in the validation period.}
	\label{fig: crpss_map}
\end{figure}

Figure \ref{fig: crpss_map} shows the color coded highest CRPSS values across all lead times over SEMOS, at all considered stations. For 99\% of the stations, one of the methods DAR-SEMOS, DAR-GARCH-SEMOS or SAR-SEMOS performs better than SEMOS in terms of CRPSS. Furthermore, for about 91\% of the stations, SAR-SEMOS yields the highest CRPSS over SEMOS, followed by DAR-GARCH-SEMOS (8\%) and DAR-SEMOS (1\%). At some stations, even CRPSS improvements over SEMOS of up to 5\% are possible, especially in mountainous areas.

\begin{figure}[h!]
	\begin{center}
		\includegraphics[scale = 0.6]{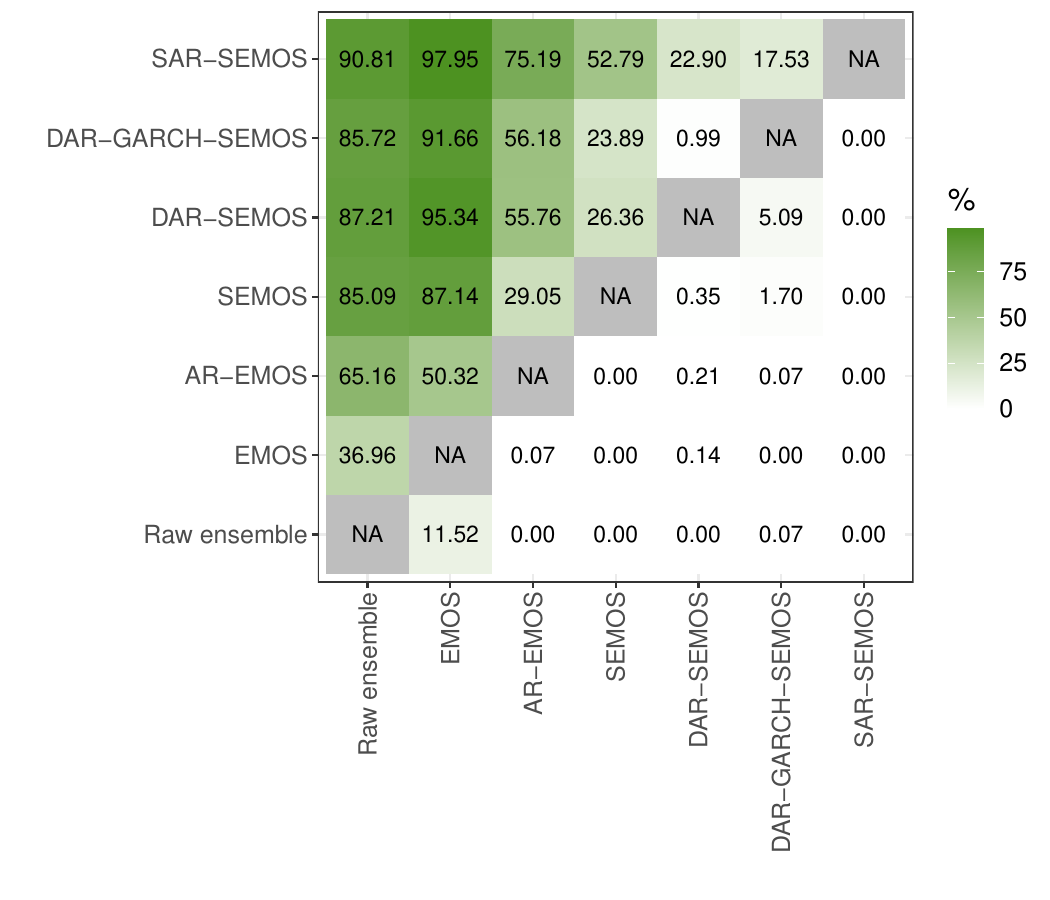}
	\end{center}	
	\caption{Percentage of pair-wise Diebold-Mariano (DM) tests with respect to lead time-station combinations indicating statistically significant CRPS differences after applying the Benjamini-Hochberg procedure with significance level $\alpha=0.05$.}
	\label{fig: sig_table}
\end{figure}

To conclude, we have a look at the statistical significance of the differences in the predictive performance with respect to CRPS among the methods in Figure \ref{fig: sig_table}. The $(i, j)$-entry in the $i$-th row and $j$-th column represents the percentage of tests over all stations and lead times, where the null hypothesis of equal predictive performance of the corresponding one-sided DM test is rejected in favor of the model in the $i$-th row when compared to the model in the $j$-th column. The difference between the sum of the $(i,j)$-th and the $(j,i)$-th entry to 100\% is the percentage where the score differences are not significant. All postprocessing methods yield a substantial portion of cases with significant CRPS improvements over the raw ensemble, while EMOS (35.96\%) and AR-EMOS (65.16\%) yield the lowest, and SAR-SEMOS (90.81\%) the highest portion of significant cases, followed by the DAR-SEMOS (87.21\%), DAR-GARCH-SEMOS (85.72\%) and SEMOS (85.09\%) models. Furthermore it is remarkable that SAR-SEMOS, DAR-GARCH-SEMOS and SEMOS yield significantly lower CRPS values than EMOS in over 85\% of all lead time-station combinations. Last but not least, it should be outlined that SAR-SEMOS significantly outperforms SEMOS in terms of CRPS in over the half of all lead time-stations combinations, while DAR-SEMOS and DAR-GARCH-SEMOS do so in around a quarter of the cases. 

\section{Conclusion and outlook}
\label{sec: Conclusion and outlook}

This work presents further time series motivated adaptions and extensions of the AR-EMOS model by \textcites{Moeller2016, Moeller2019}, and is accompanied by the \texttt{R}-package \texttt{tsEMOS}. Starting from the smooth EMOS model which takes account of seasonality in the location and scale parameter of the Gaussian distribution via finite Fourier series, this model is extended with respect to the autoregressive behavior in the forecasts errors by DAR-SEMOS. To take additionally account of autoregression in the variance, the DAR-GARCH-SEMOS model is introduced. 
Modeling autoregressive behavior in the location and scale parameter simultaneously using the standardized errors yields to the SAR-SEMOS model. 

All suggested modifications are able to successfully postprocess forecasts of arbitrary lead times and yield to a better performance in comparison to the benchmark methods EMOS and AR-EMOS for nearly each lead time. Furthermore, our modifications significantly outperform the benchmark methods in most of the lead time-station cases. Due to the simultaneous modeling of the autoregression in mean and standard deviation, SAR-SEMOS yields to the highest improvement over all methods at all stations and forecast horizons. 

As the PIT histograms in Figure \ref{fig: pit_hist} are not completely uniform, they suggest that there might still be a deficit in the proposed time series extensions of EMOS. This may be caused by a lack of important predictor variables or too inflexible marginal distributions. Therefore, we plan to investigate the latter by allowing more adjustable distribution functions, as e.g. the skew Gaussian distribution or skew Student-$t$ distribution. 
Furthermore, we plan to incorporate the autoregressive behavior of the (standardized) errors into machine learning techniques, such as e.g. gradient-boosted EMOS \parencite{Messner2017}, to join the power of statistical methods as well as machine learning methods. Table \ref{tab: significance residuals2} indicates, that SAR-SEMOS might profit from a GARCH extension as well. Last but not least, as our suggested methods are in general not restricted to any distribution assumption, adaptation to the postprocessing of other weather quantities, e.g. wind speed or precipitation is on top of our agenda. 

While we focused in this work on the statistical postprocessing of single stations, our approaches could be also extended in the spatial direction. A Gaussian random field model as proposed by \textcite{Scheuerer2013a} or a trend surface model as suggested by \textcite{Benth2012b} could be utilized to interpolate the estimated parameters of the time series based EMOS models to unobserved locations. Alternatively, the Gaussian Markov random field approach proposed by \textcite{Moeller2016b} could be applied to the (standardized) errors using the integrated nested Laplace approximation (INLA; \cite{Rue2005, Lindgren2011}) framework to account for autoregressive behavior as well. These approaches allow to be locally adaptive, while additionally taking account of spatial relationships. Furthermore, copula methods, such as the ensemble copula coupling (ECC) by \textcite{Schefzik2013}, the Gaussian copula by \textcite{Moeller2013} or vine copula based methods, such as proposed by \textcite{Jobst2023f} could be investigated to retain the spatial dependencies. Taking account of spatial autocorrelations by employing, e.g. a conditional autoregressive (CAR; \cite{Besag1974}) model might further improve local and global prediction performance. Last but not least, inter-variable or temporal dependencies could be further analyzed for which the previously mentioned copula methods could be a good starting point. 

\section*{Acknowledgements}
We are grateful to the European Centre for Medium-Range Weather Forecasts (ECMWF) and the German Weather Service (DWD) for providing forecasts and observation data, respectively. Furthermore, the authors acknowledge support of the research by Deutsche Forschungsgemeinschaft (DFG) Grant Number 395388010, and by the Hungarian National Research, Development and Innovation Office under Grant Number NN125679. Annette M\"oller acknowledges support by the Helmholtz Association's pilot project ``Uncertainty Quantification'' and by Deutsche Forschungsgemeinschaft (DFG) Grant Number 520017589.



\newpage
\addcontentsline{toc}{section}{References}
\thispagestyle{plain}
\clearpage

\printbibliography


\end{document}